\documentstyle[12pt]{article}

\begin{document}

\begin{center}

{\large \bf Gravitational waves in a stringlike fluid
cosmology} \\

\vspace{2cm}

J\'ulio C\'esar Fabris\footnote{e-mail:fabris@cce.ufes.br} \\
\vspace{0.4cm}
{\it Departamento de F\'{\i}sica, Universidade Federal do Esp\'{\i}rito
Santo,} \\
\vspace{0.1cm}
{\it Vit\'oria CEP 29060-900-Esp\'{\i}rito Santo. Brazil.} \\
\vspace{1cm}
S\'ergio Vitorino de Borba Gon\c{c}alves\footnote{e-mail:svbg@if.uff.br}
\\
\vspace{0.4cm}
{\it Instituto de F\'{\i}sica,} \\
\vspace{0.1cm}
{\it Universidade Federal Fluminense,} \\
\vspace{0.1cm}
{\it Niter\'oi CEP 24210-340-Rio de Janeiro. Brazil.} \\

\end{center}
\vspace{0.4cm}

\centerline{\bf Abstract}

\vspace{0.4cm}

{\small The coupling of a stringlike fluid with ordinary matter and
gravity may lead to a closed Universe with the dynamic of an open one.
This can provide an alternative solution for the age and horizon problems.
A study of density perturbations of the stringlike fluid indicates the
existence
of instabilities
in the small wavelength limit when it is employed a hydrodynamic
approach. Here, we extend this study to gravitational waves, where the
hydrodynamical approach plays a less important role,
and we argue that traces of the existence of this fluid must
be present in the anisotropies of the cosmic background radiation.}

\vspace{1cm}

Pacs numbers: 98.80.Cq, 04.50.+h

\vspace{0.5cm}

keywords: cosmology, cosmic microwave background.
\vspace{2cm}

\section{Introduction}
Exotic kinds of matter has been considered in many different contexts
in cosmology.
When the exotic matter has an equation of state of the type
$p = - \frac{1}{3}\rho$ ( what is currently called $K$-fluid ), it is
possible
to obtain a closed Universe ( $k = 1$ ) having the dynamics of an
open Universe\cite{kolb,toumbas}.
This fact leads to two interesting consequences:
since the Universe is closed, there is a solution for the
horizon problem without the employement of an inflationary phase;
moreover, if we are now in a phase dominated by this exotic fluid,
the scale factor behaves as $a \propto t$, and the Universe is
older in comparison with the prediction of the Standard Cosmological
Model, by a factor $\frac{3}{2}$. Since the evaluation of the age
of the Universe coming from the observational measure of the
Hubble parameter is dangerously
near the evaluated age of globular clusters, such an older Universe
can have attractive features. These models can be freely called
a "coasting" Universe, with a stringlike fluid that dominates
asymptotically the dynamic.
\par
In preceding works, we have studied the behaviour of density perturbations
in an Universe dominated by this exotic fluid\cite{julio,sergio}
using a hydrodynamic approach. We found
that there exist strong instabilities in this fluid
in the small wavelength limit that can
be reflected in the anisotropy of cosmic background radiation deduced
from density perturbations. This phenomenon occurs if we consider the
matter content of the Universe being just this exotic fluid or if it
is in presence of ordinary matter, but with no direct interaction between
them. This contradicts some previous claims that
these fluid would have no consequence at this level\cite{stelmach}.
\par
Here, we extend that study to gravitational
waves since they can also impinge traces in the anisotropy
of cosmic microwave background. The study of gravitational waves in this
scenario has many motivations.
The knowledge of the evolution of tensorial modes in an expanding
Universe is an interesting subject by itself. But there exists some
others physical motivations.
Our previous study\cite{julio,sergio} is in some sense limited, since it
consider a
hydrodynamical model only, and not fundamental fields, as it should be
done if we are more rigorous\cite{steinhardt}.
On the other hand, in the equation of gravitational waves the matter
content is not
present directly, but only through the behaviour of the scale factor.
This is not case for density perturbations, where the type of matter
we consider is determinant in the behaviour of the perturbed quantities.
So, even if we employ here again a hydrodynamic approach,
in order to have analytical expressions,
our results are caractheristics of
models where the scale factor behaves asymptotically 
as $a \propto t$, that is, dominated by the $K$-fluid, and in this sense
our final conclusions can be extended to more realistic models.
\par
The relative contribution
with respect to density perturbations is still controverse
\cite{grishchuk,jerome}. However, the study of the polarization of
the CMB photons may permit to separate the contributions of
gravitational waves and density perturbations, even though this is yet
out of observational limit\cite{melchiorri}. In the present work we will
indicate qualitatively how
the presence of that exotic fluid must be reflected in the anisotropies
of the CMB radiation, paying attention to the contribution of the
gravitational waves.

\section{Background solutions and perturbed equations}

We consider here a model where ordinary matter is coupled to an exotic
fluid, which we generally call {\it stringlike fluid}
or {\it K}-fluid. They do not interact directly between themselves, but
only
through the geometry. The Einstein's equations can be written as
\begin{eqnarray}
R_{\mu\nu} - \frac{1}{2}g_{\mu\nu}R &=& \kappa T^{(m)}_{\mu\nu} + \kappa
T^{(s)}_{\mu\nu} \quad , \\
{T^{(m)\mu\nu}}_{;\mu} &=& 0 \quad ,\\
{T^{(s)\mu\nu}}_{;\mu} &=& 0 \quad .
\end{eqnarray}
In the above expressions, $(m)$ designates ordinary matter, while
$(s)$ designates the stringlike fluid. Both fluid can be written
in a perfect fluid form, $T^{\mu\nu} = (\rho + p)u^\mu u^\nu -
pg^{\mu\nu}$. Inserting the Robertson-Walker metric
$ds^2 = dt^2 - a(t)^2\biggr(\frac{dr^2}{1 + kr^2} + r^2d\theta^2
+ r^2\sin^2\theta d\phi^2\biggl)$, we obtain the following equations
of motion,
\begin{eqnarray}
3(\frac{\dot a}{a})^2 + \frac{3k}{a^2} &=& 8\pi G(\rho_m + \rho_s) \quad
,\label{b1}\\
2\frac{\ddot a}{a} + (\frac{\dot a}{a})^2 + \frac{k}{a^2} &=& 
\frac{8\pi G}{3}(\rho_s - 3\alpha\rho_m)\quad ,\label{b2}\\
\dot\rho_m + 3\frac{\dot a}{a}(1 + \alpha)\rho_m &=& 0 \quad ,
\label{b3}\\
\dot\rho_s + 2\frac{\dot a}{a}\rho_s &=& 0 \quad . \label{b4}
\end{eqnarray}
In these expression $k$ is the curvature of the spacelike section,
$\rho_m$ is the energy density of the ordinary matter,
$\rho_s$ is the energy density of the stringlike fluid, and $p_m =
\alpha\rho_m$. Since $\rho_s \propto a^{-2}$, we can define in equation
(\ref{b1})
an effective curvature term
that can be positive, negative or zero even if we fix $k = 1$.
The resulting equation can be written
as
\begin{equation}
\frac{\dot a^2}{a^2} - \frac{\gamma}{a^2} = \frac{\lambda}{a^{3(1 +
\alpha)}}
\quad ,
\end{equation}
where $\gamma = \frac{8\pi G}{3}\rho_{0s} - k$ and
$\lambda = \frac{8\pi G}{3}\rho_{0m}$.  
The solutions, in terms of $\gamma$ and $\lambda$, expressed in terms
of the conformal time $\eta$, defined by $dt = ad\eta$, are,
\begin{itemize}
\item $\alpha = - 1$ (vacuum energy)
\begin{equation}
a = - \sqrt{\frac{\gamma}{\lambda}}\frac{1}{\sinh\sqrt{\gamma}
\eta} \quad , \quad - \infty < \eta \leq 0 \quad ;
\end{equation}
\item $\alpha = 0$ (incoherent matter):
\begin{equation}
a = \frac{\lambda}{\gamma}{\sinh^2\frac{\sqrt{\gamma}}{2}\eta} 
\quad , \quad 0 \leq \eta < \infty \quad ;
\end{equation}
\item $\alpha = \frac{1}{3}$ (radiation):
\begin{equation}
a = \sqrt{\frac{\lambda}{\gamma}}\sinh\sqrt{\gamma}\eta \quad 
, \quad 0 \leq \eta < \infty \quad;
\end{equation}
\item $\alpha = 1$
(stiff matter):
\begin{equation}
a = \biggr(\frac{\lambda}{\gamma}\biggl)^{\frac{1}{4}}\sqrt{\sinh(2
\sqrt{\gamma}\eta)} \quad , \quad 0 \leq \eta < \infty \quad.
\end{equation}
\end{itemize}
In order to perform a perturbative study of these equations, we return
back to the original
field equations putting $\tilde g_{\mu\nu} = g_{\mu\nu} + h_{\mu\nu}$,
where $g_{\mu\nu}$ is the background solution and $h_{\mu\nu}$ is
a small perturbation around it. We
impose the synchronous coordinate condition $h_{\mu0} = 0$.
We retain the tensorial modes only; the spatial behaviour of the perturbed
quantities are expressed
through normal modes
such that $\nabla^2Q_{ij} = - n^2Q_{ij}$.
We obtain the following perturbed equation governing the evolution of
gravitational waves:
\begin{equation}
\label{egw}
\ddot h_{ij} - \frac{\dot a}{a}\dot h_{ij} + \biggr[\frac{n^2}{a^2} -
2\frac{\ddot a}{a}\biggl]h_{ij} = 0 \quad .
\end{equation}
The background solutions are of the form $a = a(r\eta)$, where $r$ is a
constant.
Rewritting this equation in terms of a new conformal time,
$r\eta = \theta$,  and defining $h_{ij} \propto
h(\eta)Q_{ij}$, $Q_{ij}$ being the eigenfunction in the spatial
section with constant curvature defined before, the equation
(\ref{egw}) becomes,
\begin{equation}
\label{gw}
h'' - 2\frac{a'}{a}h' + \biggr[\tilde n^2 - 2\biggr(\frac{a''}{a}
- \frac{a'^2}{a^2}\biggl)\biggl]h = 0 \quad ,
\end{equation}
where $\tilde n^2 = \frac{1}{r^2}(n^2 + 2k)$ and the primes mean now
derivatives with respect to $\theta$.
\par
After inserting the background solutions, the equation (\ref{gw})
can be in general rewritted in terms of a hypergeometric equation.
Its final solution for different phases of the
evolution of the Universe are:
\par
\begin{itemize}
\item Vacuum energy phase ($\alpha = - 1$):
\begin{eqnarray}
h_1 &=& \sqrt{x^2 - 1}\biggr[\frac{1+x}{2}\biggl]^{- 2 + \sqrt{1 - \tilde
n^2}}\times
\nonumber\\
{_2F_1}(2 - \sqrt{1 - \tilde n^2}, \frac{1}{2} - \sqrt{1 - \tilde n^2},
1 - 2\sqrt{1 - \tilde n^2}, \frac{2}{1 + x})\quad ,\\
\nonumber\\
h_2 &=& \sqrt{x^2 - 1}\biggr[\frac{1+x}{2}\biggl]^{- 2 - \sqrt{1 - \tilde
n^2}}\times
\nonumber\\
{_2F_1}(\frac{1}{2} + \sqrt{1 - \tilde n^2}, 2 + \sqrt{1 - \tilde n^2},
1 + 2\sqrt{1 - \tilde n^2}, \frac{2}{1 + x})\quad ,\\
\nonumber\\
\tilde n^2 = \frac{1}{\gamma}(n^2 + 2k)
\quad , \quad x = \cosh\sqrt{\gamma}\eta \quad ;\nonumber
\end{eqnarray}
\item Radiative phase ($\alpha = \frac{1}{3}$):
\begin{equation}
h = \exp{(\pm(\sqrt{1 - \tilde n^2})\eta)}\sinh\eta \quad ,\quad
\tilde n^2 = \frac{1}{\gamma}(n^2 + k)
\end{equation}
\item Material phase ($\alpha = 0$):
\begin{eqnarray}
h_1 = \sqrt{x^2-1}\biggr[\frac{x+1}{2}\biggl]^{1 + \sqrt{4-\tilde n^2}}
\times \nonumber\\
{_2F_1}(-1-\sqrt{4-\tilde n^2},\frac{1}{2}-\sqrt{4-\tilde n^2},
1-2\sqrt{4-\tilde n^2},\frac{2}{1+x}) \quad ,\\
\nonumber\\
\nonumber\\
h_2 = \sqrt{x^2-1}\biggr[\frac{x+1}{2}\biggl]^{1-\sqrt{4-\tilde
n^2}}\times
\nonumber\\
{_2F_1}(\frac{1}{2}+\sqrt{4-\tilde n^2},-1+\sqrt{4-\tilde n^2},
1+2\sqrt{4-\tilde n^2},\frac{2}{1+x}) \quad ,\\
x = \cosh(\frac{\sqrt{\gamma}}{2}\eta) \quad , \quad \tilde n^2 =
\frac{4}{\gamma}(n^2 + k) \quad ;
\end{eqnarray}
\item Stiff matter ($\alpha = 1$):
\begin{eqnarray}
h_1 = \sqrt{x^2-1}\biggr[\frac{x+1}{2}\biggl]^{\frac{- 1 + \sqrt{1-4\tilde
n^2}}{2}}
\times \nonumber\\
{_2F_1}(\frac{1-\sqrt{1-4\tilde n^2}}{2},\frac{1-\sqrt{1-4\tilde n^2}}{2},
1-\sqrt{1-4\tilde n^2},\frac{2}{1+x}) \quad ,\\
\nonumber\\
h_2 = \sqrt{x^2-1}\biggr[\frac{x+1}{2}\biggl]^{\frac{-1-\sqrt{1-4\tilde
n^2}}{2}}\times
\nonumber\\
{_2F_1}(\frac{1+\sqrt{1-4\tilde n^2}}{2},\frac{1+\sqrt{1-4\tilde n^2}}{2},
1+\sqrt{1-4\tilde n^2},\frac{2}{1+x}) \quad ,\\
x = \cosh(2\sqrt{\gamma}\eta) \quad , \quad \tilde n^2 =
\frac{1}{4\gamma}\quad .
\nonumber
\end{eqnarray}
\end{itemize}
In these expressions, ${_2F_1}$ is a hypergeometric function.
We may compare the above results with the one fluid case with
$\alpha = - \frac{1}{3}$ and $k = 0$. In this case, equation (\ref{egw})
takes the form,
\begin{equation}
\ddot h - 2\frac{\dot h}{t} + n^2h = 0 \quad ,
\end{equation}
with the solution
\begin{equation}
h \propto t^{1 \pm \sqrt{1 - n^2}} \quad .
\end{equation}
\par
These solutions exhibit, for small $n^2$, growing and decreasing modes.
There is a critical value for $n^2$ over which the perturbations oscillate
with increasing amplitude.
\section{Analysis of the results}
The model discussed here represents a closed Universe
with the dynamics of an open one.
The solutions for gravitational waves are formally the same as those
we can find in an expanding Universe with ordinary matter and
$k = - 1$. 
How we can distinguish, from the analysis of gravitational waves, the open
model
from the "coasting" one? The main point is the expression for
$\tilde n^2$. Four our model,
$\tilde n^2 = \frac{1}{r^2}(n^2 + 2)$, where $r^2 = \gamma$ (for
$\alpha = -1,\frac{1}{3}$), $\frac{\gamma}{4}$ (for $\alpha = 0$) and
$4\gamma$ (for $\alpha = 1$), while of an open expanding Universe
$\tilde n = n^2 - 2$. Looking at the perturbed solutions found before,
we note that for an open Universe there is a critical value for
$n$ above which the perturbations exhibit an oscillatory behaviour, and
under which there is growing and decreasing modes.
On the other hand, for the "coasting
Universe" this behaviour occurs only for some ranges of value of
$\gamma$ while for values out of this range there is only
oscillatory behaviour. For example, in the case $\alpha = 0$,
there is only oscillatory modes for $\gamma < 2$, while for
$\gamma > 2$ we find essentially the same behaviour as in
the open Universe. So, a "coasting" and an open Universe can
become hardly indistinguishable for an extremelly high density
of the stringlike fluid, but they are clearly different if
$\gamma$ is near one.
\par
This different behaviours must lead to a spectrum of gravitational
waves perturbations completly different. The complete analysis of
this problem is out of the scope of this letter, since it touchs many
others questions like the initial spectrum of the perturbations and
its eventual quantum mechanical origin. But, we can make some qualitative
considerations.
\par
The spectrum can be obtained from the
distortion on the CMBR, comming from the integrated Sachs-Wolf
effect,
\begin{equation}
\frac{\delta T}{T} = \frac{1}{2}\int h'_{ij}e^ie^jd\eta 
\end{equation}
where $T$ is the temperature in a given direction characterized by
the unitary vector $e^i$.
The main observable quantity is the two point coorelation function
$\biggr<\frac{\delta T(e^i_2)}{T}\frac{\delta T(e^i_1)}{T}\biggl>$ which
can
be expand into multipolar coefficients through the expression
\begin{equation}
\biggr< \frac{\delta T(e^i_2)}{T}\frac{\delta T(e^i_1)}{T}\biggl> =
\sum_{l=2}^\infty c_l P_l(\cos\theta)
\end{equation}
where $\theta$ is the angle between two directions of observation
$\cos\theta = e^i_1e^i_2$.
The specific expression for the coefficients $c_l$ depends if $k = 1, 0,
-1$
\cite{caldwell} and also on the mechanism that generate the perturbations
\cite{grisha}.
\par
In any case, however, the expression for the coefficients $c_l$ depends
on the square of the classical solutions for $h$ found before. Hence,
the behaviour of $h$ is, of course, essential in the determination of the
observed quantities. We remark that for small values
of the multipole expansion of the two point correlation function,
the main contribution comes from the lower order excitations, i.e.,
$n \rightarrow 0$. We can see that, for $\gamma = 1$ and $n^2 = 0$,
an open Universe exhibits a growing mode, while the "coasting"
Universe leads to gravitational perturbations that oscillates with
increasing amplitude.
\par

A detailed analysis of the behaviour of the coefficients $c_l$
may distinguish between the many scenarios describing an expanding
Universe.
Even if the relative contribution of density perturbation is more
important than gravitational waves,
the detection of the polarization of the CMB photons may indicate
the specific contribution of the different modes, refining the
trial of cosmological scenario.
Hence, it is possible that a "coasting" Universe could be
distinguished from an expanding open Universe.
We hope to present this analysis in the future.
\vspace{1cm}

{\bf Acknowledgements:} 
It is a pleasure to thank Jos\'e Pl\'{\i}nio Baptista and
Ant\^onio Brasil Batista for many discussions during the
elaboration of this work.
We thank also CNPq (Brazil) for financial support.

\end{document}